\documentclass[aps,prl,showpacs]{revtex4}
\usepackage{amsfonts,amssymb,amsmath,latexsym,times} 
\usepackage[dvips]{graphics}

\begin{document}

[Phys. Rev. Lett. {\bf 99}, 014101 (2007)]

\title{Leading Pollicott-Ruelle Resonances and Transport in Area-Preserving Maps}

\author{Roberto Venegeroles}
\email{rveneger@fma.if.usp.br}
\affiliation{Instituto de F\'{\i}sica,
Universidade de S\~ao Paulo, C.P. 66318, 05315-970 S\~ao Paulo, S.P., Brazil}

\date{\today}

\begin{abstract}

The leading Pollicott-Ruelle resonance is calculated analytically for a general class of two-dimensional area-preserving maps. Its wave number dependence determines the normal transport coefficients. In particular, a general exact formula for the diffusion coefficient $D$ is derived without any high stochasticity approximation and a new effect emerges: The angular evolution can induce fast or slow modes of diffusion even in the high stochasticity regime. The behavior of $D$ is examined for three particular cases: (i) the standard map, (ii) a sawtooth map, and (iii) a Harper map as an example of a map with nonlinear rotation number. Numerical simulations support this formula.

\end{abstract}

\maketitle

Diffusion is a paradigm of deterministic chaos and its study is not new, dating back to Chirikov \cite{Chirikov}. Its existence in Hamiltonian systems has been extensively established using a variety of approaches \cite{Rechester,Abarbanel,Cary,McKay,Dana,Hatori}. However, it was not clear that a satisfactory transport theory could be properly formulated. In order to understand deterministic diffusion, nonequilibrium statistical mechanics was suitably combined with dynamical system theory \cite{Gaspard,Dorfman}. In this modern formulation, the stochastic properties of chaotic systems can be determined by the spectral properties of the Perron-Frobenius operator $U$. One of the most important properties is the exponential relaxation to the thermodynamic equilibrium, explained in great detail at the microscopic level. The relaxation rates $\gamma_{m}$, known as Pollicott-Ruelle (PR) resonances \cite{Pollicott,Ruelle}, are related to the poles $z_{m}$ of the matrix elements of the resolvent $R(z)=(z-U)^{-1}$ as $z_{m}=e^{\,\gamma_{m}}$. These resonances are located inside the unit circle in the complex $z$ plane, whereas the spectrum of $U$ is confined to the unit circle because of unitarity \cite{Hasegawa2}. Furthermore, the wave number dependence of the leading PR resonance determines the normal diffusion coefficient for spatially periodic systems \cite{Cvitanovic,Tasaki}. These results are rigorous only for hyperbolic systems, though they have been confirmed in the high stochasticity approximation for some mixed systems such as the kicked rotor (standard map) \cite{Khodas}, the kicked top \cite{Weber} and the perturbed cat map \cite{Blum}. The PR resonances are essential not only in classical dynamics but also in quantum dynamics. Recently, a microwave billiard experiment demonstrated a deep connection between quantum properties and classical diffusion through the spectral autocorrelation function \cite{Pance}.

In this Letter, the leading PR resonance will be calculated analytically for the general class of two-dimensional area-preserving maps
\begin{eqnarray} 
\begin{array}{l} I_{n+1} = I_{n}+K\,f(\theta_{n})\,,\\ 
                 \theta_{n+1} = \theta_{n}+c\,\alpha(I_{n+1})\qquad \mbox{mod}\, 2\pi,
\end{array} \label{map} 
\end{eqnarray}
defined on the cylinder $-\pi\leq\theta<\pi$, $-\infty<I<\infty$. Here $f(\theta)$ is the impulse function, $\alpha(I)=\alpha(I+2\pi r)$ is the rotation number, $c$ and $r$ are real parameters, and $K$ is the stochasticity parameter. This map is commonly called the {\it radial twist map} \cite{Lichtenberg} periodic in action (the nonperiodic case can be considered in the limit $r\rightarrow\infty$). Although considerable theoretical development in the study of diffusion has been achieved for the linear rotation number (LRN) case $c\,\alpha(I)\equiv I$ \cite{Rechester,Abarbanel,Cary,McKay,Dana}, many physically realistic systems are best described just by the nonlinear cases. Such maps have been extensively used in various areas of physics, especially in celestial mechanics \cite{celestial}, plasma and fluid physics \cite{plasma}, and astrophysics and accelerator devices \cite{accelerator,Lichtenberg}. However, the normal transport properties of such maps have not been studied previously \cite{Hatori}.

The analysis of the map (\ref{map}) is best carried out in Fourier space. The Fourier expansion of distribution function at the $n$th time, denoted by $\rho_{n}$, is given by
\begin{eqnarray}
\rho_{n}(I,\theta)=\sum_{m}\int dq\,e^{i(m\theta+qI)}a_{n}(m,q)\,.\label{distribution}\\\nonumber
\end{eqnarray}
The moments can be found from the Fourier amplitudes via $\langle I^{\,p}\rangle_{n}=(2\pi)^{2}\,[(i\,\partial_{q})^{p}\,a_{n}(q)]_{q=0}$, where $a_{n}(q)\equiv a_{n}(0,q)$. The discrete time evolution of the probability density $\rho$ is governed by the Perron-Frobenius operator $U$ defined by $\rho_{n+1}(I,\theta)=U\rho_{n}(I,\theta)$. The matrix representation of $U$ may be considered as the conditional probability density for the transition of the initial state $(I',\theta')$ to a final state $(I,\theta)$ in a one time step, ruled by (\ref{map}). The law of evolution of the Fourier coefficients will be given by
\begin{eqnarray}
a_{n}(m,q)=\sum_{m'}\int dq'\,\mathcal{A}_{m}(r,c,q'-q)\,\mathcal{J}_{m-m'}(-Kq')\,a_{n-1}(m',q')\,,\label{evolution}
\end{eqnarray}
where $a_{0}(m,q)=(2\pi)^{-2}\exp[-i(m\theta_{0}+qI_{0})]$. The Fourier decompositions of the $\alpha(I)$ and $f(\theta)$ functions are
\begin{eqnarray}
\mathcal{A}_{m}(r,c,x)&=&\sum_{l}\delta(lr^{-1}-x)\,\mathcal{G}_{l}(r,mc)\,,\label{Afunction}\\
\mathcal{G}_{l}(r,x)&=&\frac{1}{2\pi}\int d\theta\,\exp\{-i[x\alpha(r\theta)-l\theta]\}\,,\label{Gfunction}\\
\mathcal{J}_{m}(x)&=&\frac{1}{2\pi}\int\,d\theta\,\exp\{-i[m\theta-xf(\theta)]\}\,.\label{Jfunction}
\end{eqnarray}
If the rotation number $\alpha(I)$ is an odd function, then $\mathcal{G}_{l}(r,x)$ is a real function and $\mathcal{G}_{\pm |l|}(r,x)=\mathcal{G}_{\left|l\right|}(r,\pm\,x)$. The integral function $\mathcal{J}_{m}(x)$ assumes the following series expansions $\mathcal{J}_{m}(x)=\delta_{m,0}+\sum_{n=1}^{\infty}c_{m,n}\,x^{n}$, whose coefficients are given by
\begin{eqnarray}
c_{m,n}=\frac{1}{2\pi}\frac{i^{n}}{n!}\int\,d\theta\,f^{n}(\theta)\,e^{-im\theta}\,.
\label{coeficients}
\end{eqnarray}
Note that, if $f(-\theta)=-f(\theta)$, the coefficients $c_{m,n}$ are real for all $\left\{m,n\right\}$ and $\mathcal{J}_{-m}(x)=\mathcal{J}_{m}(-x)$.

Let us consider the decomposition method of the resolvent $R(z)$ based on the projection operator techniques utilized in \cite{Hasegawa1}. The law of evolution $(\ref{evolution})$ can be written as $a_{n}(m,q)=U^{n}a_{0}(m,q)$ where $U^{n}$ is given by the following identity $\oint_{C} dz R(z)z^{n}=2\pi i U^{n}$, where the spectrum of $U$ is located inside or on the unit circle $C$ around the origin in complex $z$-plane. The contour of integration is then a circle lying just outside the unit circle. We then introduce a mutually ortoghonal projection operator $P=\left|q,0\right\rangle\left\langle q,0\right|$, which picks out this {\it relevant} state from the resolvent and its complement $Q=1-P$, which projects on the {\it irrelevant} states. In order to calculate the diffusion coefficient $D=\lim_{n\rightarrow\infty}\,(2n)^{-1}\left\langle(I-I_{0})^{2}\right\rangle_{n}$ we can decompose the projection of the resolvent $PR(z)$ into two parts: $PR(z)=PR(z)P+PR(z)Q$. The last part can be neglected because $a_{0}(m\neq0,q)\propto e^{-im\theta_{0}}$, whose expected value disappear at random initial conditions on $[-\pi,\pi)$. Hence, the relevant law of evolution of the Fourier amplitudes, omitting initial angular fluctuations, assumes the following form:
\begin{equation}
a_{n}(q)=\frac{1}{2\pi i}\oint_{C}dz\,\frac{z^{n}}{z-\sum_{j=0}^{\infty}z^{-j}\Psi_{j}(q)}\,a_{0}(q),\label{relevantevolution}
\end{equation}
where the memory functions $\Psi_{j}(q)$ obtained for the system (\ref{map}) are given by
\begin{subequations}
\begin{equation}
\Psi_{0}(q)=\mathcal{J}_{0}(-Kq)\,,\label{psi0}
\end{equation}
\begin{equation}
\Psi_{1}(q)=\sum_{m}\mathcal{J}_{-m}(-Kq)\mathcal{J}_{m}(-Kq)\,\mathcal{G}_{0}(r,mc)\,,\label{psi1}
\end{equation}
\begin{equation}
\Psi_{j\geq2}(q)=\sum_{\{m\}}\sum_{\{\lambda\}^{\dag}}\mathcal{J}_{-m_{1}}(-Kq)\,\mathcal{J}_{m_{j}}(-Kq)\,\mathcal{G}_{\lambda_{1}}(r,m_{1}c)\prod_{i=2}^{j}\mathcal{G}_{\lambda_{i}}(r,m_{i}c)\mathcal{J}_{m_{i-1}-m_{i}}\left[-K\left(q+r^{-1}\sum^{i-1}_{k=1}\lambda_{k}\right)\right]\label{psij}.
\end{equation}
\end{subequations}
Hereafter, the following convention will be used: Wave numbers denoted by {\it Roman indices} can take only {\it non-zero} integer values, whereas wave numbers denoted by {\it Greek indices} can take {\it all} integer values, including zero. For each fixed $j$, the sets of wave numbers are defined by $\{m\}=\{m_{1},\ldots,m_{j}\}$ and $\{\lambda\}^{\dag}=\{\lambda_{1},\ldots,\lambda_{j}\}$, where the superscript denotes the restriction $\sum_{i=1}^{j}\lambda_{i}=0$.

For usual physical situations (assumed here) we have $c_{\,0,1}\propto\int d\theta f(\theta)\equiv0$ \cite{Moser}. In this case, $\Psi_{0}(q\rightarrow0)=1+\mathcal{O}(q^{2})$. In the general case, we have $\Psi_{j}(q\rightarrow0)=\mathcal{O}(q^{2})$ for $j\geq1$. The integral (\ref{relevantevolution}) can be solved by the method of residues truncating the series at $j=N$ and after taking the limit $N\rightarrow\infty$. The trivial resonance $z=1$ is related to the equilibrium state found for $m=m'=q=0$. The nontrivial leading resonance can be evaluated by the well-known Newton-Raphson iterative method beginning with $z_{0}=1$ and converging to $z_{\infty}=\sum^{\infty}_{j=0}\Psi_{j}(q)+\mathcal{O}(q^{4})$. In the limit $q\rightarrow0$, this resonance will dominate the integral in the asymptotic limit $n\rightarrow\infty$. Thus, the evolution of the relevant Fourier coefficients can be written as $a_{n}(q)=\exp[n\gamma(q)]\,a_{0}(q)$, where the leading PR resonance is given by
\begin{equation}
\gamma(q)=\ln\sum^{\infty}_{j=0}\Psi_{j}(q)+\mathcal{O}(q^{4})\,.\label{LPR}
\end{equation}
From (\ref{LPR}) the diffusion coefficient can be calculated as $D=-(1/2)\left[\partial^{2}_{q}\,\sum^{\infty}_{j=0}\Psi_{j}(q)\right]_{q=0}$. Applying this expression to the memory functions (\ref{psi0})-(\ref{psij}), the general exact diffusion coefficient formula will be given by
\begin{equation}
\frac{D}{D_{ql}}=1+2\sum_{m=1}^{\infty}\sigma_{m,m}\,\mbox{Re}[\mathcal{G}_{0}(r,mc)]+\sum^{\infty}_{j=2}\sum_{\{m\}}\sum_{\{\lambda\}^{\dag}}\,\sigma_{m_{1},m_{j}}\,\mathcal{G}_{\lambda_{1}}(r,m_{1}c)\prod_{i=2}^{j}\mathcal{G}_{\lambda_{i}}(r,m_{i}c)\,\mathcal{J}_{m_{i-1}-m_{i}}\left(-\frac{K}{r}\sum^{i-1}_{k=1}\lambda_{k}\right)
\label{diffusionformula}
\end{equation}
where $D_{ql}=-c_{\,0,2}\,K^{2}$ is the quasilinear diffusion coefficient and $\sigma_{m,m'}=(c_{-m,1}\,c_{m',1})/c_{\,0,2}$. The diffusion formula (\ref{diffusionformula}) assumes a more simple form for the LRN case (where $I$ can be replaced by $I$ mod $2\pi$), yielding $\mathcal{G}_{\lambda}(1,x)=\delta_{\lambda,x}$ and
\begin{equation}
\frac{D_{LRN}}{D_{ql}}=1+\sum_{j=2}^{\infty}\sum_{\{m\}^{\dag}}\sigma_{m_{1},m_{j}}\prod^{j}_{i=2}\mathcal{J}_{m_{i-1}-m_{i}}\left(-K\sum^{i-1}_{k=1}m_{k}\right).\label{diffusionLRT}
\end{equation}

As a check for this theory, we can first calculate $D_{LRN}$ explicitly for two cases: (i) the well-known standard map ($sm$) as an example of a mixed system and (ii) a sawtooth map $(sw)$ as a example of a hyperbolic system in a certain parameter regime. In case (i), we have $f(\theta)=\sin(\theta)$, and, hence, $\mathcal{J}_{m}(x)$ is the Bessel function of the first kind $J_{m}(x)$, $D_{ql}=K^{2}/4$, and $\sigma_{m,m'}=(\pm\delta_{m,\pm1})(\pm\delta_{m',\pm1})$. The resultant expression for $D_{sm}$ is very similar to (\ref{diffusionLRT}). The first terms of the expansion coincide with the Rechester, Rosenbluth and White results \cite{Rechester}: $D_{sm}/D_{ql}=1-2J_{2}(K)+2J_{2}^{2}(K)+\ldots$. In case (ii), we have $f(\theta)=\theta$; hence, $\mathcal{J}_{m}(x)=\frac{\sin[\pi(m-x)]}{\pi(m-x)}$, $D_{ql}=K^{2}\pi^{2}/6$, and $\sigma_{m,m'}=\frac{6}{\pi^{2}}\frac{(-1)^{m-m'}}{mm'}$. The sawtooth map is hyperbolic when $|K+2|>2$ \cite{Arnold}. Finally, we also consider a third case where $f(x)=\alpha(x)=\sin(x)$, known as the {\it Harper map} ($Hm$), as an example of a map with a nonlinear rotation number. The resultant expression for $D_{Hm}$ is very similar to (\ref{diffusionformula}), where $\mathcal{G}_{\lambda}(1,x)=J_{\lambda}(x)$ and other terms follow case (i). The analytical results on the three cases are compared with numerical calculations of $D/D_{ql}$ in Figs. 1(a)-1(c). Despite the accelerator modes, whose kinetics properties are anomalous \cite{Zaslavsky}, all theoretical results are in excellent agreement with the numerical simulations.

A question of interest that arises here is the oscillatory character of the diffusion coefficient for maps with a periodic rotation number (including the LRN case), in contrast to the fast asymptotic behavior exhibited by maps with a nonperiodic rotation number (see, for example, \cite{Hatori}). The nonperiodic case can be considered by applying the limit $r\rightarrow\infty$. For such a case, we have $\lambda r^{-1}\rightarrow s$, $r^{-1}\sum_{\lambda}\rightarrow\int ds$, and $r\,\mathcal{G}_{\lambda}(r,x)\rightarrow\mathcal{G}(s,x)$ is the $s$-Fourier transform of $e^{-ix\alpha(I)}$. For cases where $\mathcal{G}(s,x)$ is well defined, the limit $r\rightarrow\infty$ produces high oscillatory integrals resulting in $D\rightarrow D_{ql}$ without any oscillation. In the case of a standard map, Chirikov \cite{Chirikov} conjectured that the oscillatory aspect of the diffusion curve was an effect of the ``islands of stability'', but a satisfactory explanation for the oscillations has not been given yet \cite{Veneg}.

Returning to equation (\ref{diffusionformula}) we can note that, in the limit of high stochasticity parameter $K$, the diffusion coefficient does not necessarily converge to the quasilinear value in the nonlinear rotation number cases. The standard argument in this respect is the so-called {\it random phase approximation} \cite{Chirikov,Lichtenberg}. The intuitive idea is that, for large values of $K$, the phases $\theta_{n}(I,\theta)$ oscillate so fast that they become uncorrelated from $\theta$. In order to verify this effect we can take the limit $K\rightarrow\infty$ of (\ref{diffusionformula}) by setting $\lambda_{i}=0$ for all $i$ to avoid terms of order $\mathcal{O}(K^{-1/2})$. Once $|\mathcal{G}_{0}(r,mc)|<1$ for $m\neq0$, the asymptotic difusion becomes a geometric sum whose result is 
\begin{equation}
\lim_{K\rightarrow\infty}\frac{D}{D_{ql}}=1+\sum_{m\neq0}\sigma_{m,m}\frac{\mathcal{G}_{0}(r,mc)}{1-\mathcal{G}_{0}(r,mc)}.
\label{DDql}
\end{equation}
The rate (\ref{DDql}) diverges at $c=0$, creating a kind of accelerator mode. Indeed, a direct calculation through Eq.(\ref{map}) shows that $D/D_{ql}$ diverges as $n$ in this case for all $K\neq0$. In Fig. 1(d) we consider the double sine map for $K=10^{5}$. As one can see, even in the high stochasticity regime, where the random phase approximation is expected to hold, the rate $D/D_{ql}$ oscillates between the zeros of $J_{0}(c)$. Its maximum and minimum values are ruled by zeros of $J_{1}(c)$. This strong angular memory effect is a remarkable result.

Another important question concerns the higher-order transport coefficients that play a central role in the large deviations theory. These coefficients can be obtained through the following dispersion relation:
\begin{equation}
\mathcal{D}_{2l}\equiv\lim_{n\rightarrow\infty}\frac{\left\langle(I_{n}-I_{0})^{2l}\right\rangle_{c}}{(2l!)\,n}=\frac{(-1)^{l}}{(2l)!}\,\partial_{q}^{2l}\gamma(q)|_{q=0}\,,\label{disprel}
\end{equation}
where $l\geq1$ and $\left\langle\,\,\,\right\rangle_{c}$ denotes cumulant moments \cite{McLennan}. The diffusion coefficient is defined by $D=\mathcal{D}_{2}$. The higher-order coefficients $\mathcal{D}_{2l}$ can be calculated by introducing successive corrections $\mathcal{O}(q^{2l})$ in (\ref{LPR}). If the evolution process were asymptotically truly diffusive, then the angle-averaged density wold have a Gaussian contour after a sufficiently long time. A first indication of the deviation of a density function from a Gaussian packet is given by the fourth-order Burnett coefficient $B\equiv\mathcal{D}_{4}$: If $B=0$, then the kurtosis $\kappa(x)=\left\langle x^{4}\right\rangle/\left\langle x^{2}\right\rangle^{2}$ for $x=I_{n}-I_{0}$ is equal to 3 in the limit $n\rightarrow\infty$, a result valid for a Gaussian density for all times. These aspects will be treated elsewhere \cite{preparation}.

The author thanks Professor A. Saa, Professor W.F. Wreszinski, and Professor M.S. Hussein for helpful discussions. This work was supported by CAPES.

\begin{figure}[ht]
\resizebox{1.0\linewidth}{!}{\includegraphics*{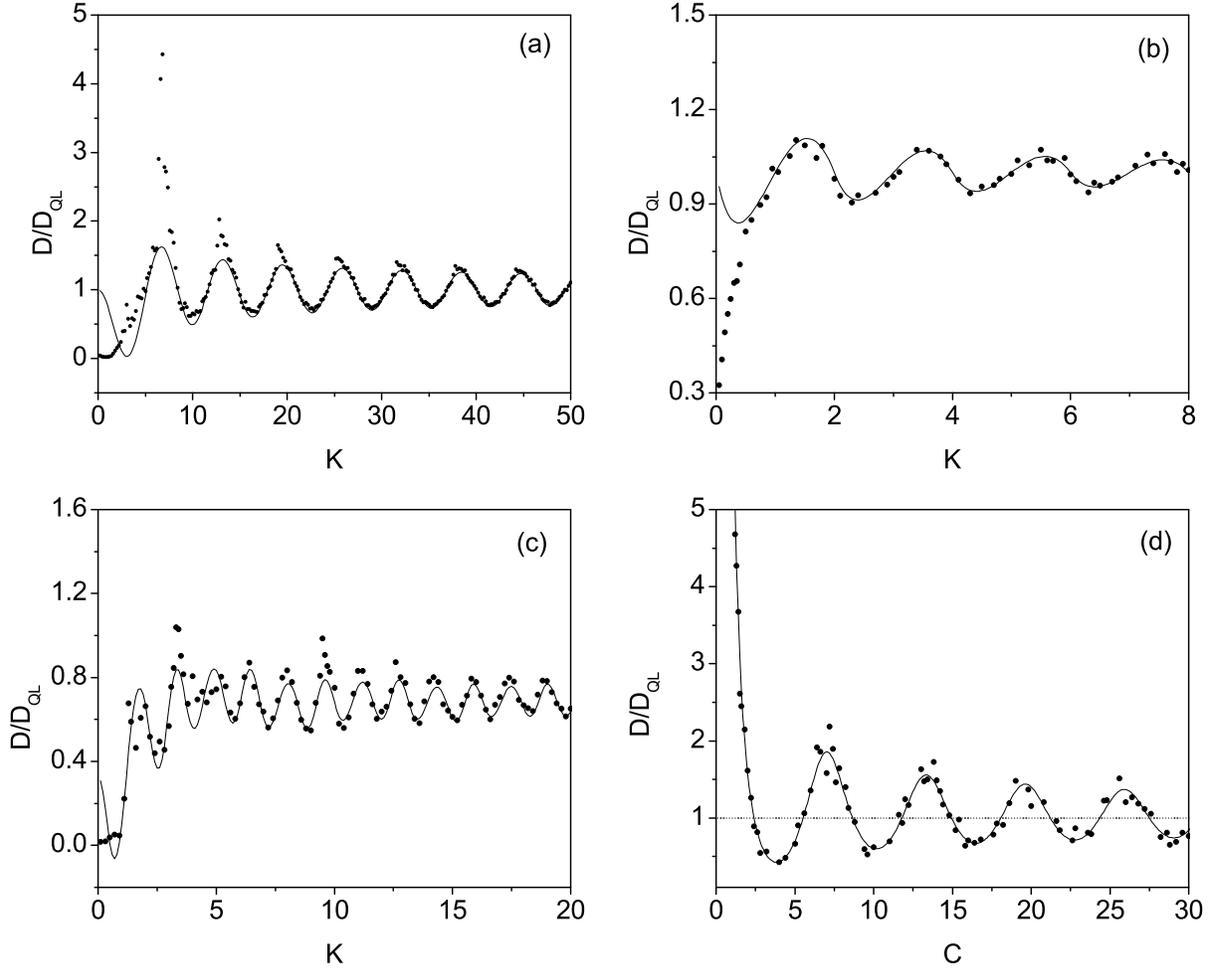}}
\caption{Theoretical diffusion coefficient rate $D/D_{ql}$ (solid lines) compared with numerical simulations calculated for $n=100$. In the cases (a), (b), and (c) we truncate the diffusion formulas (\ref{diffusionformula}) and (\ref{diffusionLRT}) at $j=2$. A better agreement for small values of $K$ requires the calculation of further memory functions. (a) Standard map. The accelerator modes give rise to spikes in the figure. (b) Sawtooth map and (c) Harper map for $c=5.5$ (with presence of accelerator modes). (d) Harper map as a function of $c$ for $K=10^{5}$. The angular evolution induces fast and slow modes of diffusion even in the high stochasticity regime. This strong angular memory effect decay as $2J_{0}(c)/[1-J_{0}(c)]$.}
\end{figure}

\end{document}